\documentclass[a4paper,fleqn,usenatbib]{mnras}

\usepackage{newtxtext,newtxmath}

\usepackage[T1]{fontenc}
\usepackage{ae,aecompl}

\usepackage{graphics,graphicx,url}
\usepackage{amssymb,amsmath}
\usepackage{bm}
\usepackage{color}

\newcommand{\beq}{\begin{equation}}
\newcommand{\eeq}{\end{equation}}

\title[The glitch-dominated rotation of PSR~J0537$-$6910]{Pulsar spin-down: the glitch-dominated rotation of PSR~J0537$-$6910}
\author[D.~Antonopoulou et al.]{D.~Antonopoulou$^{1,2}$\thanks{E-mail: antonopoulou.danai@gmail.com}, 
C. M.~Espinoza$^{3}$,  L.~Kuiper$^{4}$, and N. Andersson$^{2}$ \\
$^{1}$Nicolaus Copernicus Astronomical Center, Polish Academy of Sciences, ul. Bartycka 18, 00-716 Warsaw, Poland \\
$^{2}$Mathematical Sciences and STAG Research Centre, University of Southampton, Highfield Southampton SO17 1BJ, UK. \\
$^{3}$Departamento de F\'isica, Universidad de Santiago de Chile, Estaci\'on Central, Santiago 9170124, Chile. \\
$^{4}$SRON Netherlands Institute for Space Research, Sorbonnelaan 2, NL-3584 CA Utrecht, the Netherlands\\
}

\date{Accepted 2017 September 18. Received 2017 September 18; in original form 2017 July 25} 
 \pubyear{2017}
 
\begin{document}
\label{firstpage}
\pagerange{\pageref{firstpage}--\pageref{lastpage}}

\maketitle

\begin{abstract}

The young, fast-spinning, X-ray pulsar J0537$-$6910 displays an extreme glitch activity, with large spin-ups interrupting its decelerating rotation every $\sim$100 days. We present nearly 13 years of timing data from this pulsar, obtained with the {\it Rossi X-ray Timing Explorer}. We discovered 22 new glitches and performed a consistent analysis of all 45 glitches detected in the complete data span. Our results corroborate the previously reported strong correlation between glitch spin-up size and the time to the next glitch, a relation that has not been observed so far in any other pulsar. The spin evolution is dominated by the glitches, which occur at a rate $\sim3.5$ per year, and the post-glitch recoveries, which prevail the entire inter-glitch intervals. 
This distinctive behaviour provides invaluable insights into the physics of glitches. The observations can be explained with a multi-component model which accounts for the dynamics of the neutron superfluid present in the crust and core of neutron stars. We place limits on the moment of inertia of the component responsible for the spin-up and, ignoring differential rotation, the velocity difference it can sustain with the crust.   
Contrary to its rapid decrease between glitches, the spin-down rate increased over the 13 years, and we find the long-term braking index $n_{\rm l}=-1.22(4)$, the only negative braking index seen in a young pulsar. We briefly discuss the plausible interpretations of this result, which is in stark contrast to the predictions of standard models of pulsar spin-down. 

\end{abstract}


\section{Introduction}
\label{0537:introduction}

Much of our knowledge on neutron stars comes from pulsar timing observations, that is, tracking the rotational phase to study their spin frequency $\nu$ and its evolution. 
The slow down rate $|\dot{\nu}|$ of isolated sources depends mainly on the electromagnetic energy losses and provides, amongst other information, the test-ground for models of pulsar magnetospheres. 
Its evolution, usually expressed via the braking index $n=\nu\ddot{\nu}/\dot{\nu}^2$, can be used to explore the processes governing a neutron star's magnetic field and emission. 
This quantity, however, is hard to probe observationally: the effect of $\ddot{\nu}$ is feeble, and usually masked by timing noise. 
Furthermore, the internal dynamics (especially for relatively young pulsars) can have a strong impact on the spin behaviour. 
The most notable timing phenomenon associated with the physics of neutron star interiors is glitches: sudden and fast increases of the spin frequency.   
An increase in the spin-down rate and relaxation (over days to years) towards the pre-glitch rotational state are often seen following a glitch \citep{els+11,ymh+13}.

Both the glitch spin-up event and the subsequent slow response to it are linked to the presence of a superfluid component in the star's inner crust and core \citep{ai75,hm15}.
Neutrons in a superfluid state can be weakly coupled to the rest of the star, and thus store the required angular momentum to accelerate the crust, to which the magnetosphere and emission regions are anchored, resulting in the glitch.  
Vortex lines of quantised circulation carry the angular momentum of the superfluid and their number density defines its rotation rate. 
In equilibrium (and in the absence of spatial inhomogeneities),
 vortices are distributed in a rectilinear array and the average superfluid velocity field follows solid-body rotation, similar to that of the crust and charged components of the neutron star.  
As the pulsar slows down due to the external electromagnetic torque, the excess of vortices is removed via their motion and annihilation at the superfluid boundary. 
If such a continuous rearrangement of vortex density is prohibited (for example as a result of decreased vortex mobility due to their interaction with the nuclear lattice of the inner crust), differential rotation builds up and the spin-down of the superfluid happens in an episodic way, giving rise to glitches. 
 
Weakly coupled superfluid regions will not immediately follow the glitch spin-up, which can be very abrupt - an upper limit of $40\,\rm{s}$ has been inferred for a glitch in the Vela pulsar \citep{dcl02}. 
Such regions are then driven
 out of their equilibrium state (or further away from it) and decouple, reducing the effective moment of inertia that the external torque acts upon. 
The resulting strengthening of the spin-down rate decays on timescales characteristic of the relaxation of these regions and encodes important information about their properties, such as their temperature and dominant coupling mechanisms. 
Post-glitch relaxation can dominate the magnetospheric contribution to the evolution of the spin-down rate and be responsible for observed large ($>3$) braking indices \citep{ab06}.  
The conventional glitch model, as described above, successfully accounts for a large part of glitch phenomenology; however, several pieces of the picture remain elusive. A non-exhaustive list includes the exact trigger of vortex unpinning, the extent and location of the superfluid that participates in a glitch and the coupling strength between the various stellar components. These issues need resolving before we can extract constraints on internal properties from glitch observations and understand the effects of superfluidity on the long-term evolution of neutron stars. 

Here, we investigate the rotational history of the extraordinary pulsar PSR~J0537$-$6910 as uncovered by the $\sim13$ years of observations with the {\it Rossi X-ray Timing Explorer }({\it RXTE}). 
This pulsar holds the record as the fastest non-recycled rotation-powered pulsar known.
Moreover, with the exception of two millisecond pulsars which displayed one very small glitch each 
\citep{cb04,kjs+16}, it is the fastest pulsar observed to glitch. Its spin frequency of $\nu\simeq62\,{\rm Hz}$ is about twice that of the Crab pulsar, which is the next fastest spinning and frequently glitching neutron star. PSR~J0537$-$6910's strong spin-down rate of $\dot{\nu}=-1.992\times10^{-10}\,{\rm s^{-2}}$ is second only to that of the Crab, and its spin-down energy loss rate $\dot{E}=4.88\times10^{38}\,{\rm erg\,s^{-1}}$, the greatest known to date. 

PSR~J0537$-$6910 is a particularly interesting neutron star: it shows the highest glitch rate of any pulsar, and an atypical long-term spin-down evolution characterised by a well-defined {\it negative} braking index.  
Our analysis of the data and rotational parameters reveals a total of 45 glitches (most of which have absolute sizes amongst the largest observed in any pulsar), which display a striking regularity: the size of each spin-up strongly correlates with the time until the next one. This relation not only has interesting theoretical implications, but moreover it can be used to predict the epoch of future glitches \citep{mmw+06} and --with designated observations-- constrain the spin-up timescale and early response. 
We also present a systematic study of the glitch properties and derive limits on the basic ingredients of the glitch mechanism in a simple multifluid neutron star framework. 
Finally, we use this long dataset to explore the overall growth of the spin-down rate and its plausible physical interpretations.

\section{Observations of PSR~J0537$-$6910}
\label{0537:observations}

PSR~J0537$-$6910 was discovered by {\it RXTE} in the supernova remnant N157B \citep{mgz+98}. The remnant, located in the Large Magellanic Cloud, has a kinematic age of $<24\,\rm{kyr}$ as inferred from $H\alpha$ measurements \citep{cks+92}, and a Sedov age, estimated from its X-ray emission, of $\sim5\,\rm{kyr}$ \citep{wg98}, which is in line with the characteristic spin-down age of the pulsar ($\tau_{{\rm sd}}=4.93\,{\rm kyr}$). 
Radio pulsations have not been detected to date \citep{clj+05}, but strongly pulsed X-ray/soft $\gamma$-ray emission is observed from $0.1$ to above $50$ keV \citep[see ][ for an overview and the characteristics of the spectra and pulse profile]{kh15}. 
 
\begin{figure*}
\begin{center}
\includegraphics[width=0.98\textwidth]{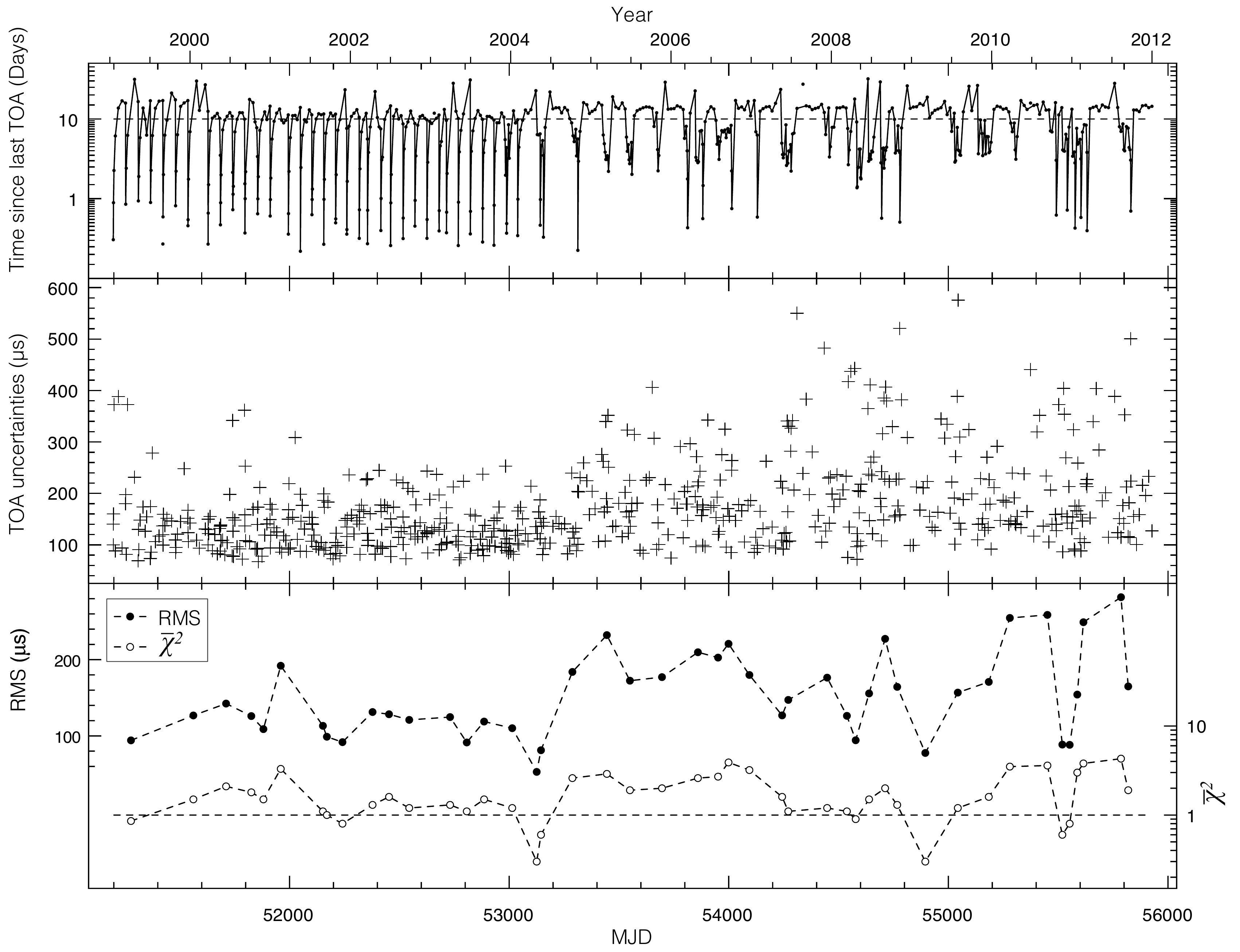}
\caption{\label{fig:TOAsTime}
Top panel: TOA separation as a function of time. Middle panel: TOA error. Lower panel: {\sc RMS} of the residuals and $\chi^2$ values of the respective fits to a timing model (see text for details on the fitting method used for each segment).}  
\end{center}
\end{figure*}

{\it RXTE} monitoring of PSR~J0537$-$6910 began on January 19, 1999 and continued until December 31, 2011. 
The general method we use to derive the time-of-arrival (TOA) of the pulses is extensively described in section 4.1 of \citet{kh09}. 
In our case, however, the template used in the correlation procedure comes from a 60 bin asymmetric Lorentzian model fit to a high-statistics pulse profile, which was obtained with the Proportional Counter Array (PCA). 
During its $\sim$16 years operational lifetime, the PCA experienced several high-voltage breakdowns in all five constituting detector units, which became more frequent as the instrument aged. The most stable unit was PCU-2, which was on almost all of the time. 
In particular, after $\sim$2006, typically one (in 50 per cent of the cases) or two (40 per cent) of the five PCUs were operational during a standard observation. This significantly reduced the sensitivity to pulsed flux detection and resulted in larger uncertainties in the reconstructed TOAs for PCA observations performed during the late stages of the {\it RXTE} mission. 
In this work, we sometimes had to combine observations which were closely spaced in time (usually by a week). 
The cadence of TOAs and their errors can be seen in the upper panels of Figure \ref{fig:TOAsTime}. Though observations were sparser in the period after 2004, typical TOA separation is less than two weeks. After 2008, TOA errors can be sometimes quite large, complicating the timing analysis.

\section{Timing analysis}
\label{0537:methods}
\subsection{Methodology} 

A preliminary reduction was carried out in order to determine time intervals (and respective TOA subsets) for which it is possible to find a phase-coherent timing solutions and to identify the epochs of candidate glitch events. All medium to large glitches ($\gtrsim10\rm{\mu Hz}$) happened at epochs when coherence was lost. Some smaller glitches, however, were found after visual inspection of the timing residuals with respect to a simple slow-down model for each interval. 

We detected a total of 45 glitches in the entire dataset, which are presented in Table \ref{tbl:glits}. 
The list comprises 24 events which occurred during the first $\sim 7$ years of data, 23 of which have been previously reported \citep{mgm+04, mmw+06}, one small glitch identified by \citet{kh15} (glitch 7 in Table \ref{tbl:glits}), and 21 new glitches in the newly examined data, after MJD 53968.

In general, the TOAs of an inter-glitch time interval were fitted with a Taylor expansion in phase $\phi$ of the form:
\begin{align}
\label{EqBasicTimingModel}
\phi(t)=&\phi_0+\nu_0\cdot(t-t_0)+\frac{\dot{\nu}_0}{2}\cdot(t-t_0)^2+\frac{\ddot{\nu}_0}{6}\cdot(t-t_0)^3
\end{align}
where $\phi_0$, $\nu_0$, $\dot{\nu}_0$ and $\ddot{\nu}_0$ are the reference phase, frequency and its first two time derivatives at epoch $t_0$.
Such a timing model provides a good fit for all time intervals except the one following the first glitch. This is due to the presence of an exponentially decaying component in the post-glitch recovery, as we discuss in section \ref{sc:glitch measuring}. All fits were done with the timing package {\sc tempo2} \citep{heesm06}.
The results are presented in Table \ref{tbl:f2s}. 

The last term of Eq. \ref{EqBasicTimingModel} is important, as it describes the gradual evolution of $\dot{\nu}$ between glitches, but its effect is significant only for the longest inter-glitch intervals. For those intervals less than $\sim40$ days long (or with exceptionally small numbers of TOAs) the quality of the fit does not significantly improve by including this term\footnote{The difference in {\sc RMS} is typically less than $10\rm{\mu s}$, the maximum was $30\rm{\mu s}$ for segment nr.41.} and resulting $\ddot{\nu}_0$ values are not accurately determined. The timing model and quoted best-fit parameters for such segments exclude that term. 

When  $\ddot{\nu}$ was detected, we used it to calculate the inter-glitch braking index $n_{{\rm ig}}$. Using the inter-glitch solutions avoids contamination of $n_{{\rm ig}}$ from the techniques used to characterise the glitches but it should be stressed that quoted values are only indicative of the overall inter-glitch spin evolution and (weakly) depend on the choice of $t_0$. Although a model with constant $\ddot{\nu}$ describes the data rather well (see last two columns in Table \ref{tbl:f2s})  with the minimum required free parameters, some departures from this linear decay of the spin-down rate are observed (asides the exponential recovery after the first glitch). For example, the resulting ``average" $|\ddot{\nu}|$ is usually smaller for longer intervals, reflecting the fact that the initial, 
 fast post-glitch recovery is slowing down at later times.


\begin{table*}
\caption{Spin parameters (ephemeris) of PSR J0537-6910, for the glitch-free observational spans. Corresponding inter-glitch braking indices $n_\text{ig}$ are calculated as $\nu\ddot{\nu}/\dot{\nu}^2$. The $1\sigma$ errors in the last quoted digit are shown between parentheses. 
  \label{tbl:f2s}}
\begin{tabular}{lcclclllccc}
\hline
\multicolumn{1}{c}{Nr.} & \multicolumn{1}{c}{Start} & \multicolumn{1}{c}{End} & \multicolumn{1}{c}{Nr. TOAs} & \multicolumn{1}{c}{Epoch} 
& \multicolumn{1}{c}{$\nu$} & \multicolumn{1}{c}{$\dot{\nu}$} & \multicolumn{1}{c}{$\ddot{\nu}$} & \multicolumn{1}{c}{$n_\text{ig}$}
& \multicolumn{1}{c}{RMS} & \multicolumn{1}{c}{$\overline{\chi}^2$}  \\

\multicolumn{1}{c}{ } & \multicolumn{1}{c}{MJD} & \multicolumn{1}{c}{MJD} & \multicolumn{1}{c}{ } & \multicolumn{1}{c}{MJD}
& \multicolumn{1}{c}{Hz} & \multicolumn{1}{c}{$10^{-15}$\,Hz\,${\rm s^{-1}}$} & \multicolumn{1}{c}{$10^{-20}$\,Hz$\,{\rm s^{-2}}$} & \multicolumn{1}{c}{ } 
& \multicolumn{1}{c}{$\mu$s} & \multicolumn{1}{c}{ }  \\
\hline
00	&	51197.1	&	51262.7	&	11	&	51229	&	62.040383761(4) &	-199227(2) &	1.0(4)  &	16(6) &	89.6	&	0.8	\\
01	&	51294.1	&	51546.7	&	33	&	51420	&	62.037138047(1) &	-199226.6(1) &	0.49(1) &	7.6(1) &	208.6	&	3.5	\\
02	&	51576.6	&	51705.2	&	18	&	51640	&	62.033378994(2) &	-199267.5(4) &	0.75(4) &	12(1) &	130.6	&	1.9	\\
03	&	51716.0	&	51817.7	&	17	&	51766	&	62.031229134(2) &	-199286(1) &	1.2(1)  &	19(1) &	155.0	&	2.4	\\
04	&	51833.8	&	51874.6	&	9	&	51854	&	62.029722720(5) &	-199294(3) &	2(1)    &	32(19) &	58.5	&	0.6	\\
05	&	51886.9	&	51954.8	&	10	&	51920	&	62.028594933(6) &	-199277(2) &	3(1)    & 	44(8) &	151.5	&	2.5	\\
06	&	51964.4	&	52144.1	&	26	&	52054	&	62.026315878(1) &	-199279.8(3) &	0.70(2) &	10.9(3) &	204.4	&	3.5	\\
07	&	52155.6	&	52165.3	&	5	&	52160	&	62.02449120(2)  &	-199090(179) &	--      &	-- &	84.6	&	1.2	\\
08	&	52175.1	&	52229.5	&	9	&	52202	&	62.023779326(3) &	-199325(3) &	1.7(5)  &	26(8) &	70.0	&	1.0	\\
09	&	52252.8	&	52367.4	&	21	&	52310	&	62.021945863(2) &	-199307.9(3) &	0.6(1)  &	10(1) &	91.1	&	0.6	\\
10	&	52389.5	&	52445.4	&	10	&	52417	&	62.020113717(9) &	-199332(4) &	1(1)    &	21(17) &	175.0	&	3.5	\\
11	&	52460.0	&	52539.0	&	14	&	52499	&	62.018715011(3) &	-199324(1) &	1.2(2)  &	18(3) &	81.1	&	0.7	\\
12	&	52551.6	&	52717.4	&	29	&	52634	&	62.016416116(1) &	-199314.2(2) &	0.72(2) &	11.2(3) &	138.4	&	1.4	\\
13	&	52745.4	&	52791.9	&	8	&	52768	&	62.014117657(3) &	-199359(5) &	--      &	-- &	83.0	&	1.0	\\
14	&	52822.8	&	52883.7	&	12	&	52853	&	62.012669329(4) &	-199365(2) &	1.8(4)  &	28(6) &	90.7	&	1.1	\\
15	&	52889.1	&	53007.2	&	20	&	52948	&	62.011047480(2) &	-199342(1) &	1.3(1)  &	21(1) &	130.9	&	1.7	\\
16	&	53019.8	&	53121.7	&	12	&	53070	&	62.008967229(1) &	-199363.5(3) &	0.95(4) &	15(1) &	49.2	&	0.2	\\
17	&	53128.1	&	53142.4	&	5	&	53135	&	62.007848725(6) &	-199393(45) &	--      &	-- &	63.5	&	0.5	\\
18	&	53147.0	&	53284.7	&	14	&	53215	&	62.006494752(2) &	-199370.1(4) &	0.87(4) &	14(1) &	102.0	&	1.1	\\
19	&	53290.9	&	53443.4	&	21	&	53367	&	62.003900833(3) &	-199377(1) &	1.1(1) &	18(1) &	252.7	&	3.5	\\
20	&	53446.8	&	53548.7	&	13	&	53497	&	62.001677516(4) &	-199399(1) &	1.7(1) &	26(2) &	184.4	&	1.8	\\
21	&	53552.0	&	53681.6	&	15	&	53616	&	61.999647307(3) &	-199395(1) &	1.1(1) &	17(1) &	164.8	&	1.9	\\
22	&	53710.6	&	53859.2	&	18	&	53784	&	61.996778414(2) &	-199397(1) &	0.83(5) &	13(1) &	188.6	&	2.0	\\
23	&	53862.1	&	53946.7	&	12	&	53904	&	61.994725551(7) &	-199426(3) &	2.2(4) &	35(6) &	239.5	&	3.6	\\
24	&	53952.7	&	53995.7	&	8	&	53974	&	61.993520808(8) &	-199349(6) &	5(2) &	81(26) &	102.7	&	1.0	\\
25	&	54002.5	&	54088.4	&	9	&	54045	&	61.992319414(3) &	-199448(3) &	-- &	-- &	248.0	&	4.6	\\
26	&	54099.3	&	54241.5	&	13	&	54170	&	61.990188310(2) &	-199429(1) &	0.73(5) &	11(1) &	143.7	&	2.2	\\
27	&	54245.0	&	54269.1	&	7	&	54255	&	61.988723956(3) &	-199360(12) &	-- &	-- &	61.6	&	0.4	\\
28	&	54273.1	&	54441.3	&	17	&	54357	&	61.986996441(2) &	-199445.5(4) &	0.80(3) &	12(1) &	191.6	&	1.3	\\
29	&	54455.0	&	54534.2	&	9	&	54494	&	61.984650498(5) &	-199471(2) &	1.6(3) &	25(5) &	141.0	&	0.9	\\
30	&	54541.8	&	54573.3	&	5	&	54557	&	61.98357186(1)  &	-199495(22) &	-- &	-- &	116.7	&	1.7	\\
31	&	54582.5	&	54637.1	&	11	&	54609	&	61.98268471(1) &	-199490(4) &	-2(1) &	-26(-21) &	82.0	&	0.7	\\
32	&	54640.3	&	54710.3	&	14	&	54675	&	61.981555144(5) &	-199469(3) &	1.8(4) &	29(7) &	207.9	&	1.8	\\
33	&	54714.3	&	54762.5	&	7	&	54738	&	61.98047594(2) &	-199475(13) &	6(3) &	98(45) &	207.5	&	2.1	\\
34	&	54770.7	&	54885.3	&	11	&	54828	&	61.978947041(2) &	-199498.4(3) &	1.1(1) &	16(1) &	65.5	&	0.2	\\
35	&	54904.1	&	55040.7	&	14	&	54972	&	61.976486188(1) &	-199477.1(4) &	0.95(3) &	15(1) &	86.9	&	0.3	\\
36	&	55044.8	&	55181.7	&	16	&	55113	&	61.974069484(3) &	-199449(1) &	2.1(1) &	33(1) &	193.4	&	1.9	\\
37	&	55185.4	&	55275.4	&	9	&	55230	&	61.972066323(5) &	-199512(1) &	0.7(2) &	11(4) &	124.7	&	1.1	\\
38	&	55284.3	&	55444.8	&	16	&	55364	&	61.969790480(4) &	-199503(1) &	0.6(1) &	9(1) &	311.6	&	4.4	\\
39	&	55457.8	&	55516.6	&	8	&	55487	&	61.967680846(7) &	-199523(2) &	2(1) &	30(11) &	102.0	&	0.9	\\
40	&	55520.7	&	55549.3	&	7	&	55535	&	61.966861019(3) &	-199528(8) &	-- &	-- &	64.8	&	0.3	\\
41	&	55562.5	&	55584.4	&	7	&	55573	&	61.966206683(7) &	-199412(20) &	-- &	-- &	109.5	&	1.3	\\
42	&	55589.3	&	55610.5	&	6	&	55599	&	61.96576392(2) &	-199574(49) &	-- &	-- &	196.9	&	4.4	\\
43	&	55619.0	&	55786.1	&	16	&	55702	&	61.964016055(4) &	-199529.2(5) &	0.88(5) &	14(1) &	276.3	&	3.7	\\
44	&	55794.7	&	55818.6	&	5	&	55806	&	61.962221550(6) &	-199428(23) &	-- &	-- &	92.3	&	0.8	\\
45	&	55823.0	&	55926.9	&	11	&	55872	&	61.961105096(5) &	-199549(1) &	1.1(2) &	16(3) &	200.4	&	2.5	\\

\hline
\end{tabular}
\end{table*}

\subsubsection{Glitches}
\label{sc:glitch measuring}
Typically, glitches are described by additional terms in Eq. \ref{EqBasicTimingModel}, that set in after the glitch epoch $t_{\rm g}$:
\begin{align}
\nonumber \phi_\mathrm{g}(t)=&\Delta\phi+\Delta\nu_\mathrm{p}\cdot(t-t_\mathrm{g})+
                    \frac{\Delta\dot{\nu}_\mathrm{p}}{2}\cdot(t-t_\mathrm{g})^2+\frac{\Delta\ddot{\nu}_\mathrm{p}}{6}\cdot(t-t_\mathrm{g})^3\\
\label{EqTimingModelGlitch}
                     &-\left(\sum_i\Delta\nu_\mathrm{d}^{(i)}\tau_\mathrm{d}^{(i)} e^{-(t-t_\mathrm{g})/\tau_\mathrm{d}^{(i)}}\right).
\end{align}
Quantities with index $\rm{p}$ describe persisting step-like (unresolved in time) changes in the rotational parameters, while an index $\rm{d}$ denotes exponentially decaying components in the post-glitch timing residuals. In the following we will drop the index $\rm{p}$ for simplicity - decaying components will still be denoted with the index $\rm{d}$. 
The term $\Delta\phi$ ensures phase coherence between the pre-glitch and post-glitch TOAs, which is lost if the glitch epoch is not known precisely. 

The determination of $t_\mathrm{g}$ is difficult and depends mainly on the TOA coverage and the size of the glitch.
With the reasonable assumption that the rotational phase is continuous across the glitch, the time of the event can be pinpointed by finding the time at which $\Delta\phi=0$.
Unfortunately, this leads to unique solutions only for relatively small glitches and/or when the observational gap around $t_\mathrm{g}$ is small. 
It was possible to use the condition $\Delta\phi=0$ to determine the glitch epoch with higher accuracy only for the six smallest glitches.
For the other glitches, $t_\mathrm{g}$ is poorly constrained and was defined as the central point between the pre-glitch and post-glitch sets of TOAs.  The error was calculated as half the distance between the two datasets, leading to uncertainties of $5$ to $10$\,days.

The glitch parameters were found by fitting the function $\phi(t)+\phi_\mathrm{g}(t)$ to the TOAs around the assumed glitch epoch and delimited by the previous and next glitches.
For all glitches but the first one, we dismiss the last term of Eq. \ref{EqTimingModelGlitch} in the model because no evidence for exponential relaxations are present. 
We allow instead for a non-zero jump in $\ddot{\nu}$, which accounts for the change in gradient of $\dot{\nu}$ with one parameter less than an exponential term. This was found sufficient to describe the post-glitch datasets. As expected, however, if the pre- or post-glitch interval is too short, $\ddot{\nu}$ will be poorly detected over that period of time, compromising the measurement of $\Delta\ddot{\nu}$. The situation gets worse when both intervals surrounding $t_\mathrm{g}$ are short. 
We based the decision of including or not these terms on the results of glitch simulations, as described in the next section.  
The outcome of the simulations offered also a way to assess the uncertainties on our derived parameters.

\subsubsection{Simulations}
\label{sec:simulations}

We performed glitch simulations in order to define a consistent technique, which would enable the recovery of glitch parameters as accurately as possible for all glitches in PSR~J0537$-$6910. 
While $\Delta\nu$ is almost always well constrained, the spin-down changes $\Delta\dot{\nu}$ are much more uncertain and sensitive to the methods and models used to handle the data, especially if the TOA coverage is poor and the error bars large. 
Additionally, although PSR~J0537$-$6910 exhibits large $\ddot{\nu}$ between glitches, many inter-glitch intervals are too short to allow its precise determination. Because of this, the presence of a change $\Delta\ddot{\nu}$ at a glitch is sometimes unclear and hard to quantify.
The inclusion, or not, of $\ddot{\nu}$ and $\Delta\ddot{\nu}$ in the glitch timing model can significantly vary the measured values of $\Delta\dot{\nu}$ and, though to a lesser extent, $\Delta\nu$.
We use simulations to examine the output of a range of timing models in order to optimise the measurement of $\Delta\nu$ and $\Delta\dot{\nu}$ for this particular pulsar.
The investigated models varied mostly on the treatment of the $\ddot{\nu}$ terms of Eqs. \ref{EqBasicTimingModel} and \ref{EqTimingModelGlitch}
\footnote{A detailed analysis of the efficacy of glitch measuring techniques, based on a big sample of simulated data aimed to represent different sources and monitoring programmes, will be presented elsewhere (Espinoza et al. in prep.).}.

Simulated sets of TOAs, following the measured rotation of PSR~J0537$-$6910, were produced with the {\sc toasim} plugins available in {\sc tempo2}.
Four different situations were identified in the real data and reproduced separately in the simulations: cases in which both pre- and post-glitch intervals are long ($\gtrsim 100$ days, which also ensures that there are more than 10 TOAs in each ``long" subset of data), cases in which only the post-glitch interval is long, cases in which only the pre-glitch interval is long and cases in which both intervals are short.
The TOA uncertainties and cadence for the simulated datasets were taken from the real data.
For that we used two representative examples of pre- and post-glitch sets for each of the four cases described above.

A glitch was introduced at an epoch drawn from a uniform probability distribution covering 20--30 days around the inferred glitch epoch of the original data.
Glitch parameters $\Delta\nu$, $\Delta\dot{\nu}$ and $\Delta\ddot{\nu}$ were randomly drawn from representative normal distributions, which were derived from preliminary measurements of these parameters for the 45 glitches.
For $\Delta\ddot{\nu}$ we considered only measurements coming from cases in which both the pre- and post-glitch intervals are long.
Exponentially decaying terms, which in general are not favoured by the data, were not included in the simulations. 

The simulated glitches were modelled by global fits, to the pre- and post-glitch data, of an underlying spin-down timing model plus the additional glitch terms (Eqs. \ref{EqBasicTimingModel} and \ref{EqTimingModelGlitch}). The best-fit parameters were then compared to the original values.
The fitting functions corresponded to all possible combinations between including or not the terms $\ddot{\nu}$ and $\Delta{\ddot{\nu}}$.
We also tested models in which $\ddot{\nu}$ was kept at a constant value, chosen as the average of the observed values for several inter-glitch TOA sets, or as the value corresponding to the longest of the pre- and post-glitch intervals.
For the cases in which both intervals are short ($<100$ days) we performed extra simulations of small glitches and tested, in addition, models with and without the $\Delta\dot{\nu}$ term. 
This was motivated by preliminary results showing that the detection of this term is hard for small glitches, which --in the case of this particular pulsar-- are always surrounded by very short datasets.

We found that the inclusion of all terms leads to the most accurate measurements of $\Delta\nu$ and $\Delta\dot{\nu}$ only when both the pre- and post-glitch intervals were long ($\gtrsim 90$ days). 
For the other three cases however, where one or both intervals are short, the real parameters are better recovered if we remove the $\Delta{\ddot{\nu}}$ term and fit for $\ddot{\nu}$, even if the latter is not well constrained (i.e. its fractional error is larger than 1). From the simulations focussed on the small glitches and short intervals we saw that even if $\Delta\dot{\nu}$ is poorly constrained, setting it to zero negatively affects the accuracy of the measured $\Delta\nu$. 

The glitch measurements of simulated data serve also as a method to probe the uncertainties of the glitch parameters. 
Usually, standard ($1\sigma$) errors from the fitting procedures are reported in the literature, although they are often underestimates of the true errors and do not account for the uncertainty in the glitch epoch. 
According to the simulations, for the chosen fitting functions used in this analysis, $1\sigma$ errors are underestimated, in general, by a factor of $\sim2$. 
Consequently, we applied this multiplying factor to all best-fit standard errors\footnote{This means that our reported best-fit errors on glitch parameters (e.g. in Table \ref{tbl:glits}) should be viewed as approximately representing a $\sim 68\%$ confidence interval.}. 

As already mentioned, when $t_\mathrm{g}$ cannot be uniquely determined it is set to the mid-point between the pre- and post-glitch datasets.
Our simulations results confirm that this is typically a reasonable choice, often with little impact on the accuracy of inferred parameters. 
Nonetheless, the errors arising from the uncertainty in $t_\mathrm{g}$ should not be ignored.
This is particularly important if the gap in the TOAs around $t_\mathrm{g}$ is long.
We quantified these uncertainties by performing two additional fits per glitch, setting $t_\mathrm{g}$ to be the epoch of the TOAs bracketing the event. 
The error was then taken as half of the difference of the measured parameters at the two boundaries and was compared to the errors of the fitting procedure with $t_\mathrm{g}$ set at the mid-point. We quote the largest of the two in Table \ref{tbl:glits}.

\subsection{Results}
\label{0537:results}

All 45 glitches were analysed according to our findings from measuring simulated glitches (section \ref{sec:simulations}). 
That is, it is best to fit for $\ddot{\nu}$ and set $\Delta\ddot{\nu}=0$ in all cases in which any, or both, of the datasets prior and after the glitch are shorter than $\sim$90 days. 
Both terms should be included otherwise, which was the case for 18 glitches in this sample. 
 The precise choice of 90 days is an empirical one, based on the outcome of the simulations. It does, however, have some physical justification as this is approximately the timescale over which the $\ddot{\nu}$ terms of the fitted equations become comparable to the errors on the higher order $\dot{\nu}$ terms. We note that when $\Delta\ddot{\nu}$ is not used and set to zero, the $\ddot{\nu}$ value obtained from the fit is contaminated by the glitch and has no physical meaning.
Meaningful measurements of $\ddot{\nu}$ were performed differently, from fits only to glitch-free intervals (Table \ref{tbl:f2s}). 
The two smallest glitches required the additional exclusion of the $\Delta\dot{\nu}$ term, as explained below. The lower panel of Figure \ref{fig:TOAsTime} displays the statistics of the fitting procedure ({\sc rms} and $\chi^2$ values of the residuals).

The glitch parameters are presented in Table \ref{tbl:glits}. Their properties are described and discussed in section \ref{0537:glitches}. 
As discussed earlier, $\ddot{\nu}$, $\Delta\dot{\nu}$ and $\Delta\ddot{\nu}$ are sometimes unconstrained (as reflected by $>1$ fractional errors) but still required in the fit in order to recover correctly the rest of the parameters. 
This is the case for 15 glitches, out of which 10 have only $\Delta\ddot{\nu}$ unconstrained. 


\begin{table*}
\caption{Parameters for the 45 glitches of PSR J0537-6910, the fitted data range (Start/End epoch) and best-fit statistics ({\sc RMS} and $\overline{\chi}^2$). Details on the methods used to characterise the glitches and quantifying the parameters' errors (last digit errors in parentheses) are presented in Section \ref{0537:methods}.   
  \label{tbl:glits}}
\begin{tabular}{lcclcccccc}
\hline
\multicolumn{1}{c}{G. Nr.} & \multicolumn{1}{c}{Start} & \multicolumn{1}{c}{End} & \multicolumn{1}{c}{G. Epoch} & \multicolumn{1}{c}{$\Delta\phi$} 
& \multicolumn{1}{c}{$\Delta\nu$} & \multicolumn{1}{c}{$\Delta\dot{\nu}$} & \multicolumn{1}{c}{$\Delta\ddot{\nu}$} 
& \multicolumn{1}{c}{RMS} & \multicolumn{1}{c}{$\overline{\chi}^2$}  \\

\multicolumn{1}{c}{ } & \multicolumn{1}{c}{MJD} & \multicolumn{1}{c}{MJD} & \multicolumn{1}{c}{MJD} & \multicolumn{1}{c}{ } 
& \multicolumn{1}{c}{$\mu$Hz} & \multicolumn{1}{c}{$10^{-15}$\,Hz$\,{\rm s^{-1}}$} & \multicolumn{1}{c}{$10^{-20}$\,Hz$\,{\rm s^{-2}}$} 
& \multicolumn{1}{c}{$\mu$s} & \multicolumn{1}{c}{ }  \\
\hline
1	&	51197.1	&	51423.0	&	51278(16)  &	0.61(3)  &	42.4(2)  &	-123(12)  &	--  &	94.2	&	0.9	\\
2	&	51423.0	&	51705.2	&	51562(15)  &	0.63(3)  &	27.9(2)  &	-148(15)  &	0.3(3)  &	126.8	&	1.5	\\
3	&	51576.6	&	51817.7	&	51711(5)  &	-0.06(2)  &	19.5(1)  &	-123(13)  &	0.5(2)  &	142.4	&	2.1	\\
4	&	51716.0	&	51874.6	&	51826(8)  &	0.13(2)  &	8.7(1)  &	-100(16)  &	--  &	126.0	&	1.8	\\
5	&	51833.8	&	51954.8	&	51881(6)  &	0.15(2)  &	8.7(1)  &	-139(44)  &	--  &	108.9	&	1.5	\\
6	&	51886.9	&	52144.1	&	51960(5)  &	1.1(2)  &	28.2(3)  &	-156(94)  &	-2(1)  &	192.0	&	3.3	\\
7	&	51997.4	&	52165.3	&	52152(1)  &	0.00(1)  &	0.15(2)  &	--  &	--  &	113.2	&	1.1	\\
8	&	51997.4	&	52229.5	&	52170(5)  &	0.68(2)  &	11.4(1)  &	-155(32)  &	1(1)  &	98.9	&	1.0	\\
9	&	52175.1	&	52367.4	&	52241(12)  &	0.37(2)  &	26.44(5)  &	-48(12)  &	--  &	91.8	&	0.8	\\
10	&	52252.8	&	52445.4	&	52378(11)  &	0.05(2)  &	10.4(1)  &	-85(16)  &	--  &	131.3	&	1.3	\\
11	&	52389.5	&	52539.0	&	52453(7)  &	-0.06(3)  &	13.52(5)  &	-76(40)  &	--  &	128.4	&	1.6	\\
12	&	52460.0	&	52717.4	&	52545(6)  &	-0.4(1)  &	26.1(1)  &	-92(40)  &	-0.4(5)  &	121.0	&	1.2	\\
13	&	52551.6	&	52791.9	&	52731(14)  &	-0.04(3)  &	9.0(2)  &	-128(12)  &	--  &	124.7	&	1.3	\\
14	&	52745.4	&	52883.7	&	52807(15)  &	-0.1(1)  &	15.8(2)  &	-125(59)  &	--  &	91.4	&	1.1	\\
15	&	52822.8	&	53007.2	&	52886(3)  &	-0.38(1)  &	14.55(2)  &	-87(9)  &	--  &	118.8	&	1.5	\\
16	&	52889.1	&	53121.7	&	53014(6)  &	0.50(1)  &	21.0(1)  &	-143(12)  &	-0.4(2)  &	110.1	&	1.2	\\
17	&	53019.8	&	53142.4	&	53125.5(1)  &	0.00(1)  &	1.0(1)  &	-83(69)  &	--  &	52.8	&	0.3	\\
18	&	53019.8	&	53284.7	&	53145(2)  &	0.29(1)  &	24.25(1)  &	-38(7)  &	-0.1(1)  &	81.2	&	0.6	\\
19	&	53152.4	&	53443.4	&	53288(3)  &	-0.18(2)  &	24.51(4)  &	-137(15)  &	0.3(2)  &	183.9	&	2.6	\\
20	&	53290.9	&	53548.7	&	53445(2)  &	-0.35(2)  &	16.09(4)  &	-174(21)  &	0.6(4)  &	232.2	&	2.9	\\
21	&	53446.8	&	53681.6	&	53550(2)  &	0.00(2)  &	19.90(4)  &	-134(23)  &	-0.6(3)  &	172.4	&	1.9	\\
22	&	53552.0	&	53859.2	&	53696(14)  &	1.0(1)  &	25.4(2)  &	-139(19)  &	-0.2(2)  &	177.1	&	2.0	\\
23	&	53710.6	&	53946.7	&	53861(1)  &	-0.14(2)  &	14.56(4)  &	-167(28)  &	1(1)  &	209.7	&	2.6	\\
24	&	53862.1	&	53995.7	&	53951.3(3)  &	0.00(3)  &	1.1(1)  &	-59(47)  &	--  &	202.7	&	2.7	\\
25	&	53952.7	&	54088.4	&	53999(3)  &	-0.19(3)  &	21.9(1)  &	-88(54)  &	--  &	220.8	&	3.9	\\
26	&	54002.5	&	54241.5	&	54094(5)  &	-0.8(1)  &	23.0(2)  &	-18(52)  &	1(1)  &	179.9	&	3.2	\\
27	&	54099.3	&	54269.1	&	54243(8)  &	0.00(1)  &	0.06(2)  &	--  &	--  &	126.9	&	1.6	\\
28	&	54245.0	&	54441.3	&	54271(2)  &	-0.27(2)  &	30.3(1)  &	-154(41)  &	--  &	147.2	&	1.1	\\
29	&	54273.1	&	54534.2	&	54448(7)  &	-0.26(2)  &	14.8(1)  &	-151(30)  &	1(1)  &	176.4	&	1.2	\\
30	&	54455.0	&	54573.3	&	54538(4)  &	0.28(2)  &	7.1(1)  &	-108(51)  &	--  &	126.3	&	1.1	\\
31	&	54541.8	&	54637.1	&	54578(5)  &	0.36(4)  &	9.1(1)  &	68(145)  &	--  &	94.3	&	0.9	\\
32	&	54582.5	&	54710.3	&	54639(2)  &	-0.39(2)  &	7.98(3)  &	-84(48)  &	--  &	155.7	&	1.5	\\
33	&	54640.3	&	54762.5	&	54712(2)  &	0.22(4)  &	6.5(1)  &	-109(58)  &	--  &	227.3	&	2.0	\\
34	&	54714.3	&	54885.3	&	54767(4)  &	-0.50(2)  &	22.4(1)  &	-112(28)  &	--  &	164.3	&	1.3	\\
35	&	54770.7	&	55040.7	&	54895(9)  &	0.31(2)  &	21.1(1)  &	-103(11)  &	-0.1(1)  &	77.6	&	0.3	\\
36	&	54904.1	&	55181.7	&	55043(2)  &	-0.09(2)  &	13.45(3)  &	-159(16)  &	1.2(2)  &	156.8	&	1.2	\\
37	&	55044.8	&	55275.4	&	55184(2)  &	-0.12(2)  &	12.94(4)  &	-223(27)  &	-1(1)  &	170.8	&	1.6	\\
38	&	55185.4	&	55444.8	&	55280(4)  &	-0.4(1)  &	34.0(2)  &	-63(58)  &	0(1)  &	254.7	&	3.5	\\
39	&	55284.3	&	55516.6	&	55451(7)  &	0.23(3)  &	10.47(4)  &	-79(15)  &	--  &	258.7	&	3.6	\\
40	&	55457.8	&	55549.3	&	55519(2)  &	-0.45(1)  &	7.58(4)  &	-87(53)  &	--  &	88.6	&	0.6	\\
41	&	55520.7	&	55584.4	&	55552(2)  &	0.0(1)  &	0.5(1)  &	323(413)  &	--  &	88.3	&	0.8	\\
42	&	55562.5	&	55610.5	&	55587(2)  &	-0.3(1)  &	5.4(1)  &	-242(823)  &	--  &	154.2	&	3.0	\\
43	&	55589.3	&	55786.1	&	55615(4)  &	0.15(4)  &	28.1(1)  &	-32(91)  &	--  &	249.1	&	3.8	\\
44	&	55619.0	&	55818.6	&	55786.06(1)  &	-0.02(3)  &	0.9(1)  &	46(39)  &	--  &	281.9	&	4.3	\\
45	&	55794.7	&	55926.9	&	55819(2)  &	-0.19(1)  &	21.4(1)  &	-181(71)  &	--  &	164.9	&	1.9	\\
\hline
\end{tabular}
\end{table*}   

A decaying term, replacing $\Delta\ddot{\nu}$, was necessary to produce featureless (flat) residuals only after the first glitch, where a small amplitude exponential recovery with timescale $\tau\sim20$ days is detected. This is the largest glitch in the data and it is followed by a much longer glitch-free interval ($\sim280$\,d) compared to all other glitches ($<180$\,d). 
While exponential recoveries could be following the other glitches too, their smaller magnitude and shorter post-glitch intervals render them undetectable. 

The parameters for the first glitch are presented in Table \ref{tbl:glit1}. 
We also perform a fit using a smaller post-glitch interval, for which the exponential term can be omitted, for consistency with the derivation of parameters for the rest of the glitches; the resulting values of this fit are the ones presented in Table \ref{tbl:glits}.   
Note that for the fit of the following (second) glitch, we use a shortened pre-glitch interval, to avoid the strong decaying phase. 

\begin{table*}
\caption{Glitch parameters for the first glitch (the largest spin-up found in this study), accounting for the presence of an exponential recovery.
  \label{tbl:glit1}}
\begin{tabular}{lcclcccccccc}
\hline
 \multicolumn{1}{c}{Start (MJD)} & \multicolumn{1}{c}{End (MJD)} & \multicolumn{1}{c}{Glitch Epoch (MJD)} & \multicolumn{1}{c}{$\Delta\phi$} 
& \multicolumn{1}{c}{$\Delta\nu$ ($\mu$Hz)} & \multicolumn{1}{c}{$\Delta\dot{\nu}$ ($10^{-15}$\,Hz$\,{\rm s^{-1}}$)} & \multicolumn{1}{c}{$\Delta\nu_\mathrm{d}$ ($\mu$Hz)} & \multicolumn{1}{c}{$\tau_\mathrm{d}$ (days)} 
& \multicolumn{1}{c}{RMS ($\mu$s)} & \multicolumn{1}{c}{$\overline{\chi}^2$}  \\

\hline
51197.1 & 51546.7 & 51278(16)  &	-0.5(1)  & 	42.3(2)  &	-70(7)  &	0.3(1)  & 	21(4)	&	104.6 & 0.9	\\
\hline
\end{tabular}
\end{table*}

For glitches 7 and 27, the two smallest glitches (with $\Delta\nu<1.0\,\mu$Hz), we were unable to constrain both $\Delta\nu$ and $\Delta\dot{\nu}$, not only due to their magnitude but also because they are followed by extremely short post-glitch intervals.
Glitch 27 is the smallest and has a pre-glitch interval with just 7 TOAs, covering 24 days. In this case, results could only be obtained from a fit to a model with $\Delta{\nu}$ alone.
Glitch 7 happened only 9.6 days before the next glitch, an interval containing only 5 TOAs.
Fitting only for a frequency change results in the value presented in Table \ref{tbl:glits}. 
This model gives $\Delta\phi=0$ in between the established pre- and post-glitch datasets, consistent with what visual examination of the data suggests.
It should be noted though that the observed departure from the pre-glitch timing model can be equally well fitted by a change in spin-down rate alone, taking place (i.e. the time at which $\Delta\phi=0$) before the last TOA of the pre-glitch dataset. 
This fit offers a very similar reduced $\chi^2$ value as the fit for $\Delta{\nu}$ alone and leads to a positive change $\Delta\dot{\nu}=(100\pm8)\times10^{-15} \,\rm{Hz/s}$, i.e. a ``timing noise'' kind of event. 

While all glitches were fitted individually, without including other glitches in the pre- or post-glitch datasets, for glitches 8 and 18 it was necessary to break this rule. 
In both cases the pre-glitch datasets contain only 5 TOAs (the fewest of any inter-glitch intervals), hence it was necessary to include TOAs from before the previous glitch to create new, longer datasets.
The parameters of the previous glitches were kept fixed to their respective best-fit values during the new fits.
Both cases concern very small events, unlikely to contaminate significantly the measurements of glitches 8 and 18, which are large ones.
We note, nonetheless, that their reported sizes are measured mainly with respect to the timing solutions valid prior to glitches 7 and 17, respectively.

\begin{figure}
\begin{center}
\includegraphics[width=0.5\textwidth]{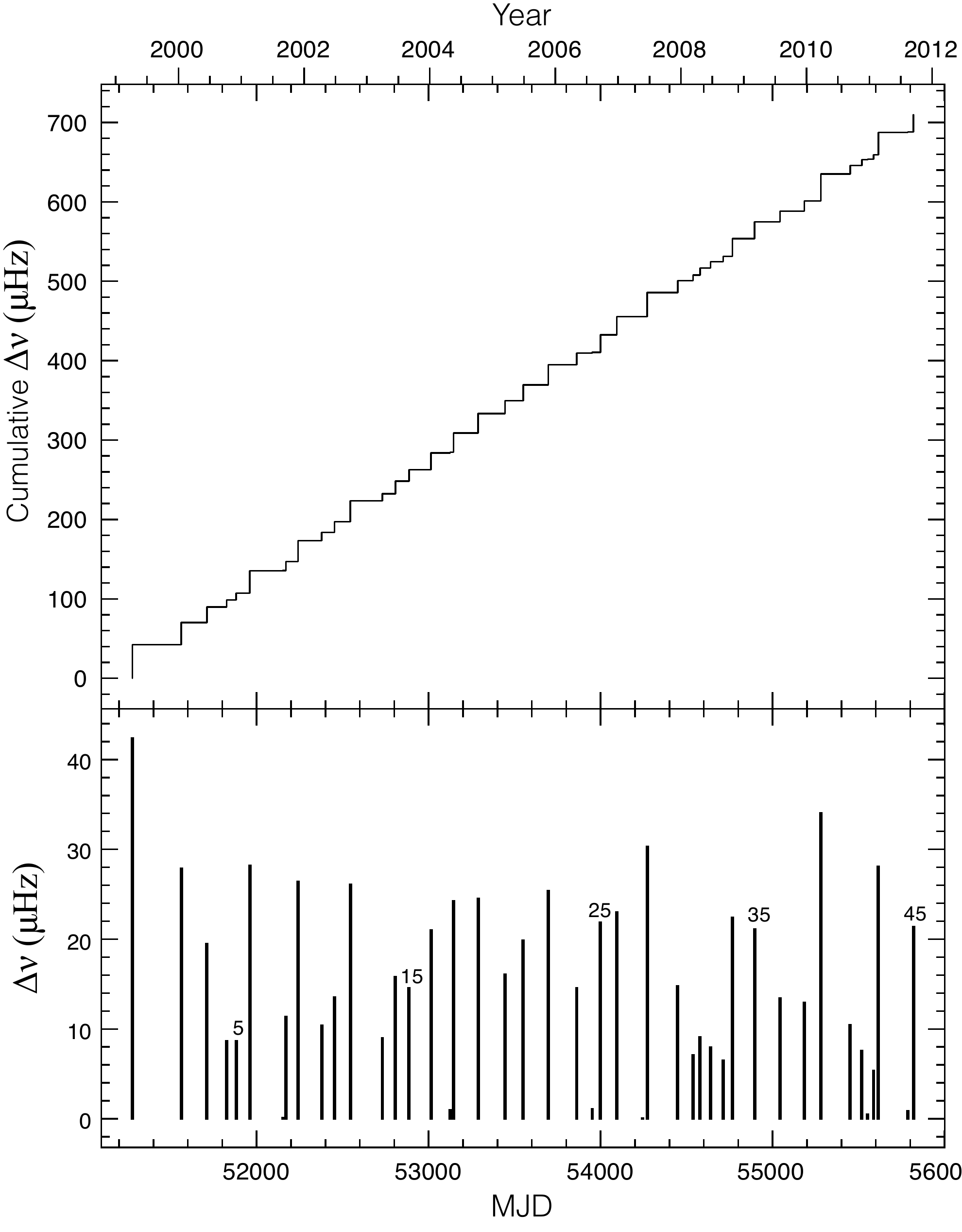}
\caption{\label{fig:glitches} Glitches in PSR~J0537$-$6910. Top panel: The cumulative increase in spin frequency over time due to the abrupt changes at glitches. The glitch activity parameter (see Section \ref{dF0}) can be extracted by a linear fit in this plot. Lower panel: The sizes and distribution of the 45 glitches in time.}  
\end{center}
\end{figure}

The sequence of glitches in time, as well as their magnitude in spin frequency\footnote{For the first glitch, the displayed size is the sum of the persisting and decaying component.} are shown in the lower panel of Figure \ref{fig:glitches}, while the upper panel displays the cumulative increase in frequency due to the glitches. 
The average inter-glitch time interval is $\sim103$ days, leading to a glitching rate of $\sim3.5$ per year, the highest observed so far in any pulsar. We do not find evidence for significant variations of the glitch activity with time.
 
Finally, the history of the spin-down rate is presented in Figure \ref{fig:F1}. As can be seen in this plot, the effect of glitches in the spin-down evolution is rather dramatic. A quasi-linear decay of $|\dot{\nu}|$ is clearly seen after each glitch (or, in the case of the first glitch, once the exponential recovery no longer dominates), implying inter-glitch braking indices well above 3; on the other hand, overall, there is a net increase in the spin-down rate over the course of the observations. This, in turn, can be associated with a negative long-term ``braking index''.   
We will discuss the physical interpretation of these results in the following.

\section{Glitch properties and interpretation}
\label{0537:glitches}

\begin{figure}
\begin{center}
\includegraphics[width=0.5\textwidth]{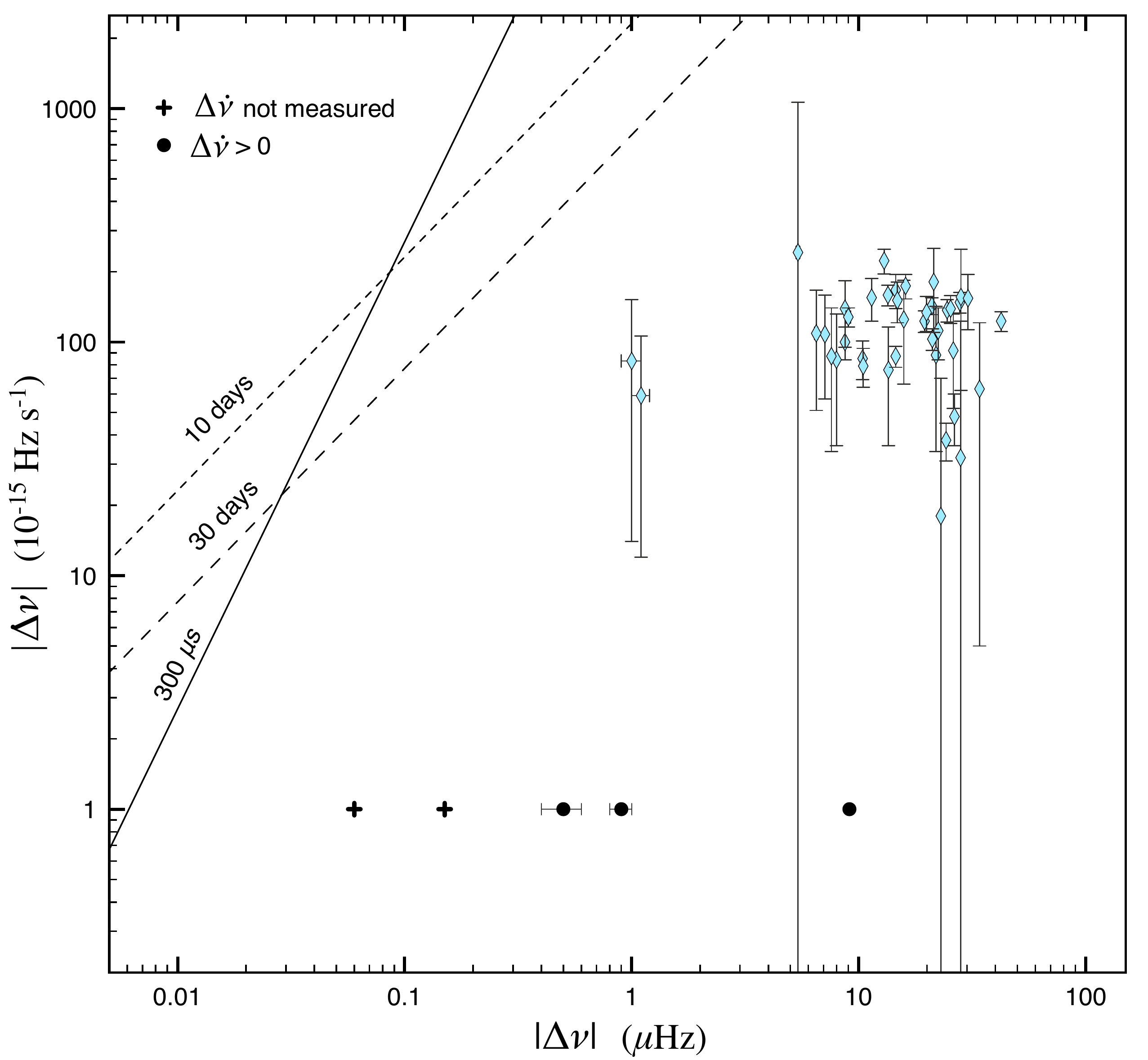}
\caption{\label{fig:limits} Conservative glitch detection limits and the changes in frequency, $\Delta\nu$, and spin-down rate, $|\Delta\dot{\nu}|$, for the 40 glitches with $\Delta\dot{\nu}<0$ (blue diamonds). Only the change $\Delta\nu$ is shown for the 3 glitches with $\Delta\dot{\nu}>0$ (black circles) and the two glitches for which $\Delta\dot{\nu}$ was not fitted for (black crosses). The solid line represents the detection limit for phase residuals or TOA error of $300\, \rm{\mu s}$, while the two dashed lines correspond to the limits due to TOA separation greater than 10 and 30 days. Glitches with ($\Delta\nu,\Delta\dot{\nu}<0$) above these lines would not have been identified in observations with such respective parameters.}  
\end{center}
\end{figure}

In the previous section we identified and parameterized, as consistently as possible, 45 glitches contained in the {\it RXTE} data of PSR~J0537$-$6910. 
Figure \ref{fig:limits} presents these glitches in a $\Delta\dot{\nu}-\Delta{\nu}$ diagram, together with detectability limits for recovering ($\Delta\dot{\nu}<0$) glitches derived according to Equation 2 in \citet{eas+14}. We use a value of $300\, \rm{\mu s}$ as the maximum between TOA error and the {\sc RMS} of the (inter-glitch) phase residuals, and two different observation cadences of 10 and 30 days. The values for these three parameters were chosen in order to obtain a very conservative estimate of the detection limits. As can be seen in figure \ref{fig:TOAsTime}, typically the TOA error and {\sc RMS} are much smaller than $300\, \rm{\mu s}$, and observations are usually separated by no more than two weeks. This gives us confidence that our glitch sample is complete above, at least, $\Delta\nu\simeq 1\,{\rm \mu Hz}$. Most likely all glitches in the examined dataset that had $\Delta\nu\gtrsim 0.3\,{\rm \mu Hz}$ have been detected, unless they had an exceptionally large $\Delta\dot{\nu}$ and occurred in the few observing periods with TOA separation $>20-30$ days. 
We will now examine the properties of the glitch population of PSR~J0537$-$6910, and discuss their implications in a simple superfluid glitch model framework.

\subsection{The frequency changes}
\label{dF0}

The spin-ups, $\Delta\nu$, span about two orders of magnitude. 
In other pulsars with such a broad range of glitch sizes, a power-law provides a good description of the distribution \citep{mpw08,eas+14}. 
In the case of PSR~J0537$-$6910 however, glitch sizes appear to be normally distributed, with a mean $\simeq15.9 \;\rm{\mu Hz}$ and standard deviation $\simeq 10.8 \;\rm{\mu Hz}$ (probability $85\%$ and $89\%$ using the Cramer von Mises or K--S test respectively). The glitch ``waiting'' times $\Delta T$ are consistent with a Weibull distribution (Cramer von Mises probability $94\%$), with shape and scale parameters $1.74$ and $118.4$ days respectively (with a mode of 72.5 days and mean 105.5 days). 

\begin{figure}
\begin{center}
\includegraphics[width=0.5\textwidth]{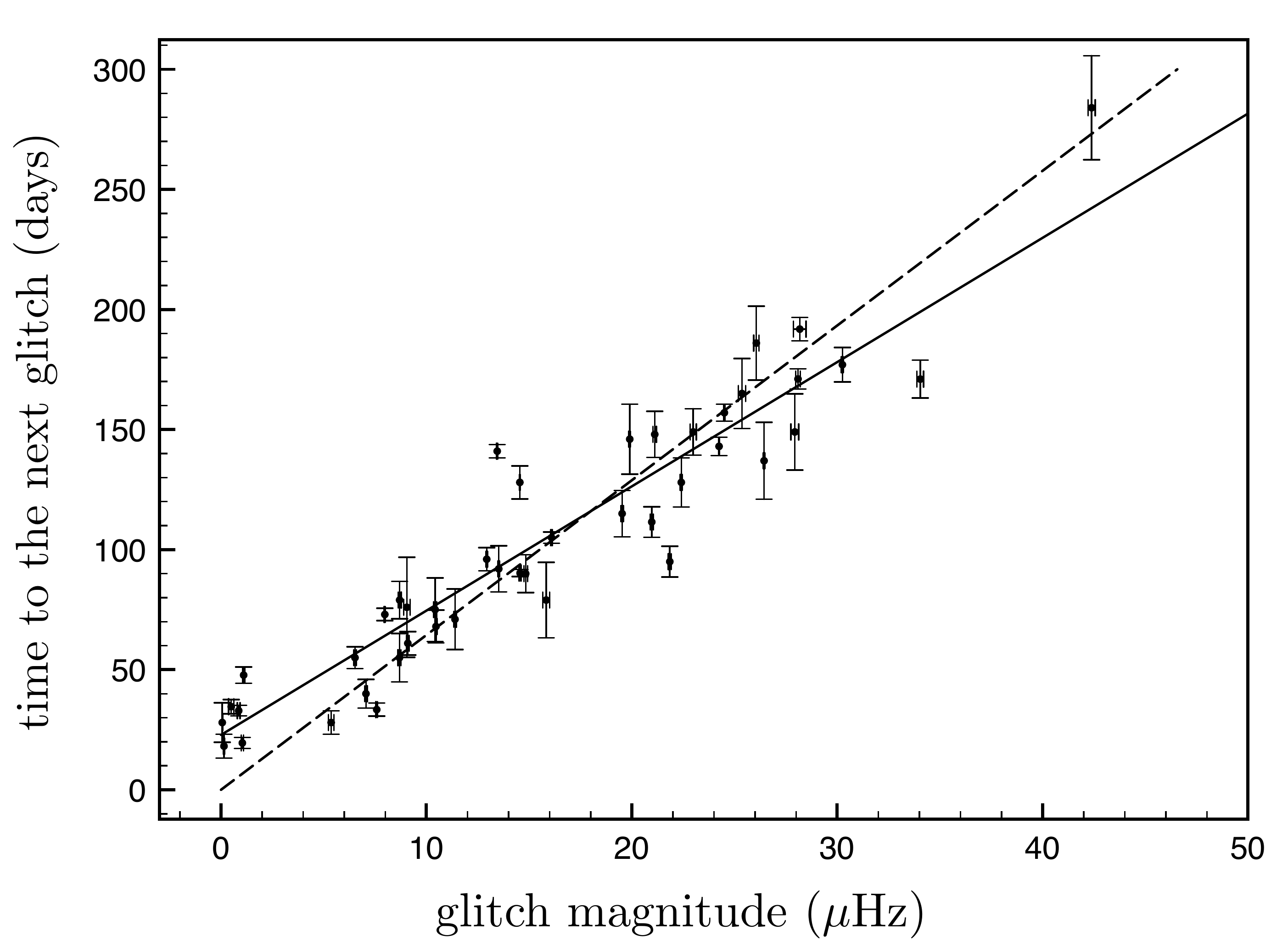}
\caption{\label{fig:df0_cor} The time interval to the next glitch versus the spin-up at a glitch. The best linear fit description of the data is $\Delta T=5.17(8)\Delta\nu+22.8(1.1)$ (solid line), while a physically motivated fit imposing the line to go through the origin returns a slope $\sim 6.4(1) \; \rm{days\,{\mu Hz}^{-1}}$ (dashed line).} 
\end{center}
\end{figure}

The regularity in glitch size and recurrence times (Figure \ref{fig:glitches}) suggests a common underlying mechanism and trigger for the large glitches in PSR~J0537$-$6910.
This is further confirmed by a correlation between the size $\Delta\nu$ of a glitch to the time $\Delta T$ until the next one, as can be seen in Figure \ref{fig:df0_cor}. 
Instead of simply using the time between glitches, however, it is more natural to compare the amount by which the star span down before the next glitch, $\dot\nu\Delta T$, with the size $\Delta\nu$ of the preceding one. These two quantities are indeed also correlated. 

For the first 23 glitches, \citet{mmw+06} report a correlation coefficient $0.94$ between $\Delta\nu$ and $\Delta T$. 
We verify the existence of a strong and significant correlation using our twofold sample (45 glitches): the linear correlation coefficient (Pearson's coefficient) for our sample is $\sim0.95$ (with probability $p_{{\rm P}}\sim3\times10^{-22}$) and the Spearman's rank correlation one is $r_{\rm sr}=0.95$ ($p_{\rm S}\sim5\times10^{-23}$). 
Almost identical results are obtained for the correlation between $\Delta\nu$ and $\dot{\nu}\Delta T$ (with the latter calculated as the integral of $\dot{\nu}(t)$ over $\Delta T$ using the inter-glitch solutions of Table \ref{tbl:f2s}). This implies that glitches are triggered when some critical threshold is reached due to the spin-down, and their sizes are related to the departure from this threshold at each event. 

There are two main candidates for the quantity that builds up to its critical value: stresses in the crust, and the rotational lag $\omega$ between the superfluid and ``normal'' stellar component. Crustquakes, however, cannot be solely responsible for the observed spin-up since such large events are expected to be far less common than the observed glitch rate. Most, if not all, of the increase in spin must be the result of angular momentum transfer from the internal superfluid. We will thus assume in the following that glitches originate from the same superfluid region that rapidly transfers angular momentum to the crust when its critical rotational lag $\omega_{\rm cr}$ is exceeded. This naturally gives rise to the observed correlation as the glitch size $\Delta\nu=\Delta\Omega_{\rm c}/2\pi$ will be proportional to the the local drop in $\omega$ (offset from $\omega_{\rm cr}$). On the other hand, the size of a glitch does not correlate with the time since the last one, as could perhaps be expected if, for example, bigger glitches were driven by a larger superfluid region that had time to reach its critical lag. 

Let us assume that neutron vortices in the glitch-driving region are strongly pinned (completely immobilised) whilst $\omega<\omega_{\rm cr}$. 
At the glitch, vortices catastrophically unpin, move (and possibly re-pin) on a much shorter timescale compared to the inter-glitch evolution. 
Therefore, the spin-down rate for this superfluid component (hereafter denoted ${\rm G}$) is zero, $\dot{\Omega}_{\rm G}=0$, between glitches.  
At the same time, the crust and all stellar components tightly coupled to it - summing up to a moment of inertia $I_{\rm c}$ - spin down at a rate $\dot\Omega_{\rm c}(t)=2\pi\dot{\nu}_{\rm c}(t)$, where $\dot{\nu}_{\rm c}(t)$ can be calculated from the inter-glitch timing model. 
The rotational lag $\omega_{\rm G}(t)=\Omega_{\rm G}-\Omega_{\rm c}(t)$ after a glitch at epoch $t_{\rm g}$ thus evolves as 
\beq
\label{lag}
\omega_{\rm G}(t>t_{\rm g})-\omega_{\rm G}(t_{\rm g})=\int -\dot\Omega_{\rm c}(t) dt \,. 
\eeq

Excess angular momentum in the component ${\rm G}$ builds at a rate $\leq I_{\rm G}|\dot\Omega_{\rm c}|$, which should be compared to its transfer rate to the observed component, $I_{\rm c}A$, where \[A=\frac{1}{T_{\rm obs}}\sum\Delta\Omega\] is the activity parameter. 
For this pulsar $A$ is very well defined due to the high glitch rate and regularity, as illustrated in Figure \ref{fig:glitches}. 
A linear fit suggests that the moment of inertia of the component ${\rm G}$, $I_{\rm G}$, accounts for $\geq0.873 \pm 0.005 \%$ of the moment of inertia that follows the spin-up, $I_{\rm c}$.
Even if the latter comprises most of the stellar moment of inertia, $I_{\rm c}\sim I_{\rm{tot}}$, and there is strong crustal entrainment, the inferred $I_{\rm G}$ can be accommodated by the inner crust for a reasonable range of neutron star masses (\citealp{c13}, see however \citealp{hea+15} for the possible core contribution).  

The spread of glitch sizes 
 could arise either from differences in the fractional moment of inertia participating in each glitch, or incomplete transfer of the excess angular momentum (the decrease in lag $|\delta\omega_{\rm G}|\neq\omega_{\rm cr}$), or both. 
Let us examine the second possibility in more detail. 
If corotation between region ${\rm G}$ and the crust is restored at a glitch ($\omega_{\rm G}(t_{\rm g})=0$), Eq. \ref{lag} becomes $\omega_{\rm G}(t>t_{\rm g}) =\Omega_{\rm c}(t_{\rm g})-\Omega_{\rm c}(t)$. 
By extrapolating the inter-glitch timing solutions to the epochs of consecutive glitches, we find that the maximum possible lag that is built up can vary even by an order of magnitude. 
This discrepancy, unlikely to be due to a large change of $\omega_{\rm cr}$ on such a short timescale, indicates that $\omega_{\rm G}$ does not always drop to zero at a glitch and the observed range of glitch sizes can be explained by differences in $\delta\omega_{\rm G}$ alone. We are thus justified to assume, for simplicity, that $I_{\rm G}$ is the same for each glitch.

Adopting $I_{\rm G}/I_{\rm c}=8.73\times10^{-3}$, an estimate for a lower limit of the critical lag follows from the size $\Delta\Omega_{\rm c}$ of the largest observed glitch. 
By angular momentum conservation, $I_{\rm G}|\Delta\Omega_{\rm G}|=I_{\rm c}\Delta\Omega_{\rm c}$, 
\beq
\delta\omega_{\rm G}=-\frac{I_{\rm c}+I_{\rm G}}{I_{\rm G}}\Delta\Omega_{\rm c}
\eeq
 where $\delta\omega_{\rm G}$ is the drop in $\omega_{\rm G}$. For $\Omega_{\rm G}\geq\Omega_{\rm c}$, $|\delta\omega_{\rm G}|$ must not be in excess of $\omega_{\rm cr}$, thus
\beq
\omega_{\rm cr}\geq\frac{I_{\rm c}+I_{\rm G}}{I_{\rm G}}\Delta\Omega_{\rm c}\simeq3\times10^{-2}\,\rm{rad\,s^{-1}}.
\eeq 
The lag increase during the longest observed inter-glitch interval (between the first and second glitch) leads to a very similar minimum for $\omega_{\rm cr}$ of $3.1\times10^{-2} \,\rm{rad\,s^{-1}}$, but relies on the uncertain extrapolation of the inter-glitch timing solution to the glitch epochs. 

If we were to relax our assumptions, the limits for the critical lag would change. 
For example, a partial coupling of region ${\rm G}$ ($\dot{\Omega}_{\rm G}<0$) would reduce the inferred $\omega_{\rm cr}$ and require a realistic treatment of the superfluid's differential rotation. On the other hand, if the region where vortices are immobilised is not disconnected from superfluid regions closer to the rotational axis that allow vortex currents, vortices could accumulate there and the real critical lag could even be higher than our estimate. 


\subsection {The spin-down rate changes}

\begin{figure*}
\begin{center}
\includegraphics[width=0.999\textwidth]{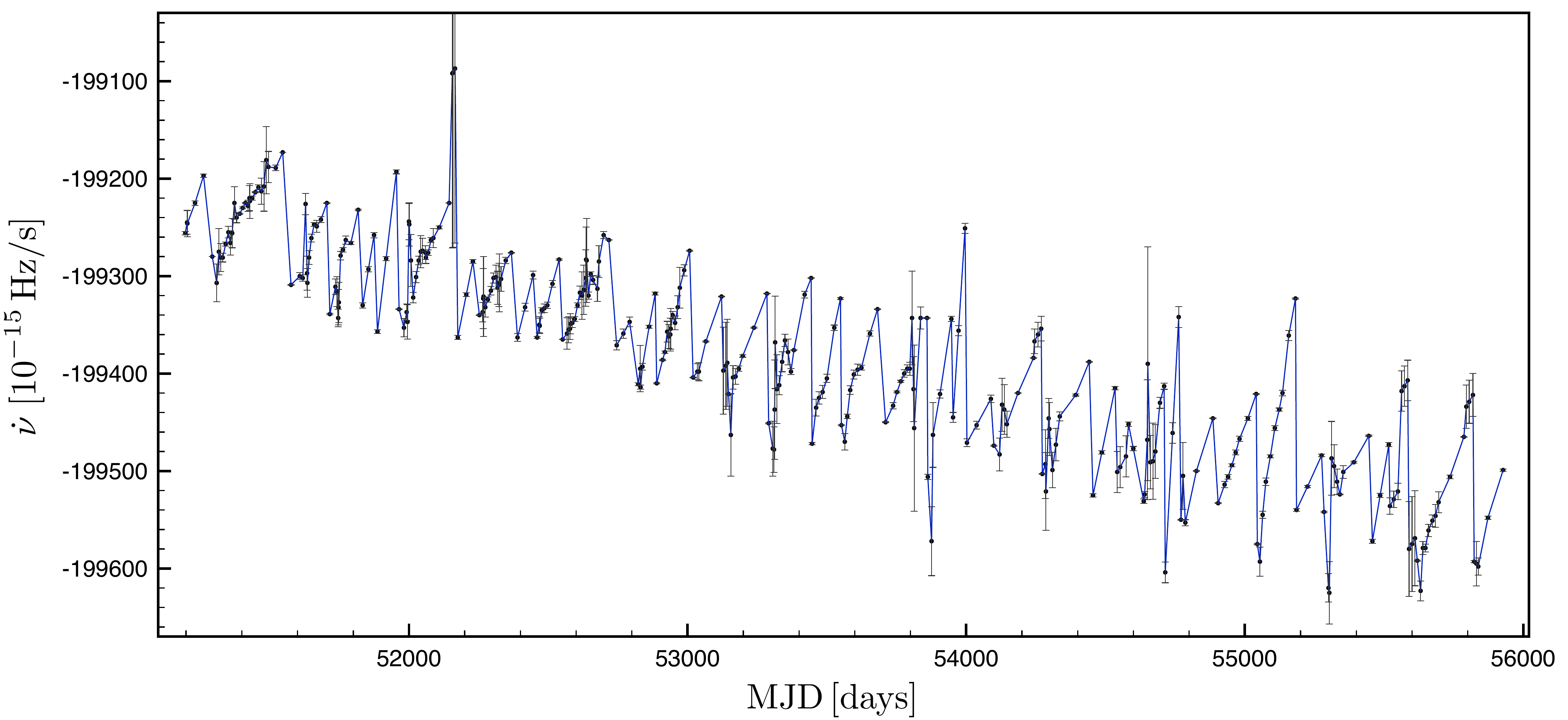}
\caption{\label{fig:F1}The spin-down rate evolution of PSR J0537$-$6910. These points have been calculated only for illustrative purposes, by fitting Eq.\ref{EqBasicTimingModel} on overlapping subsets of minimum 5 TOAs (and minimum 30 days long, when possible) -- fitting over glitch epochs was avoided. In these timing models, $\ddot{\nu}$ values were held fixed to their best-fit value for the entire inter-glitch interval (as presented in Table \ref{tbl:f2s}) or, when such a measurement was not available, to a weighted average value of $0.6 \times10^{-20} \,\rm{Hz\,s^{-2}}$. The values at the epochs of the first and last TOA of each inter-glitch interval were calculated via fits to the entire interval. }
\end{center}
\end{figure*}  

A negative change of the spin-down $\dot{\Omega}_{\rm c}=2\pi\dot{\nu}$, unresolved in time, accompanies the majority of glitches. 
The post-glitch $\dot{\nu}$ evolution is well approximated by a linear increase (decreasing magnitude of spin-down rate), until the epoch of the next glitch. 
The only exception to this nearly linear behaviour is the first (and largest) glitch, for which the inclusion of a short-term ($\sim20 \,\rm{days}$) exponential recovery in the timing model significantly improves the residuals. 
The inter-glitch evolution is characterised by a positive $\ddot{\nu}$ of the order $10^{-20}\,\rm{Hz\,s^{-2}}$, and large braking indices ($n_{\rm ig}\sim22$ on average, see Table \ref{tbl:f2s}). 

The measurements of $\Delta\dot{\nu}$ are less accurate than those of $\Delta\nu$, hindering a robust statistical analysis.
Using only those glitches (37 out of 45) that have a clear detection of $\Delta\dot{\nu}<0$, the best-fit distribution is normal with mean $\simeq-118.4\times10^{-15} \;\rm{Hz\,s^{-1}}$ and standard deviation $\simeq 44.4 \times10^{-15}\;\rm{Hz\,s^{-1}}$ (Cramer von Mises probability $94\%$). 
We do not observe a hard upper limit of $|\Delta\dot{\nu}|\simeq150\times10^{-15}\,\rm{Hz\,s^{-1}}$ as suggested by \citet{mmw+06}, but according to the above distribution the probability for $|\Delta\dot{\nu}|>200\times10^{-15}\,\rm{Hz\,s^{-1}}$ is indeed very low. 

The changes $\Delta\dot{\nu}$ for all glitches do not appear to correlate with $\Delta\nu$. Furthermore, we do not find evidence for the correlation between the time interval preceding a glitch and the change $\Delta\dot{\nu}$ which was reported by \citeauthor{mmw+06}. 
We noted though that the smallest glitches of our sample ($\Delta\nu<1\,\rm{\mu Hz}$) do not demonstrate the same behaviour as the rest, that is, none of them has a clear negative change in $\dot{\nu}$. Instead, glitches 7, 27 and 41 have $\Delta\dot{\nu}$ consistent with being zero, and the timing solution for glitch 44 gave a positive $\Delta\dot{\nu}$ but with a large fractional error ($0.85$) thus potentially also had no effect on $\dot{\nu}$. 
Assuming that these small spin-ups do not act to ``reset'' the clock for the spin-down rate changes\footnote{For example, this could be the case if the inferred spin-up was gradual, rather than abrupt as in larger glitches. Excluding these small irregularities from the glitch sample has little impact on the $\Delta\nu$ analysis and does not alter the main conclusions.}, we calculated ``waiting times'' only between glitches with $\Delta\nu>1\,\rm{\mu Hz}$ (41 events). Then a somewhat weak but significant correlation ($r_{sr}=0.57$, $p_{S}\sim2\times10^{-4}$) emerges.  

In the simplest case, the neutron star comprises of three components, $I_{\rm c}$, $I_{\rm G}$ and $I_{\rm n}$ such that the total moment of inertia is $I_{\rm tot}=I_{\rm c}+I_{\rm G}+I_{\rm n}$ and the total angular momentum $L_{\rm tot}=I_{\rm c}\Omega_{\rm c}+I_{\rm G}\Omega_{\rm G}+I_{\rm n}\Omega_{\rm n}$. 
The component $I_{\rm n}$ consists of any superfluid region that is loosely coupled to $I_{\rm c}$: if vortices are not immobilised due to pinning (as in the region ${\rm G}$), their outwards motion allows for a non-zero $|\dot{\Omega}_{\rm n}|\leq|\dot{\Omega}_{\rm c}|$, with the equality defining an ``equilibrium'' lag $\omega_{\rm s}$. 
Since $\dot{\Omega}_{\rm G}(t\neq t_{\rm g})=0$, $I_{\rm G}$ is decoupled before and after the glitch and does not contribute to the observed inter-glitch spin-down, which is described by
\beq
I_{\rm c}\dot{\Omega}_{\rm c}+I_{\rm n}\dot{\Omega}_{\rm n}=N_{\rm ext} \;\text{.}
\eeq  
For standard dipole braking, the external torque is 
\beq
\label{externaltorque}
N_{\rm ext}=-\frac{B_\perp^2 R^6_\star}{6c^3}\Omega^3_{\rm c} ,
\eeq
where $B_\perp=B_{\rm p}\sin\alpha$, $B_{\rm p}$ the polar dipole magnetic field component, $\alpha$ its angle to the rotational axis, and $R_\star$ the stellar radius. 
At any given moment, the observed magnitude of the spin-down rate is therefore bound between a maximum of $|\dot{\Omega}_{\rm c}|^{\rm max}=|N_{\rm ext}|/I_{\rm c}$, and a minimum $|\dot{\Omega}_{\rm c}|^{\rm min}=|N_{\rm ext}|/(I_{\rm c}+I_{\rm n})$ when the ``equilibrium'' lag is reached.  
When a glitch occurs, any loosely coupled superfluid region will be driven out (or further away) of its equilibrium lag, which results in the observed abrupt increase in spin-down rate.    

The change in $\dot{\Omega}_{\rm c}$ provides a lower limit for the component $I_{\rm n}$ that can decouple at the glitch. 
Assuming a torque as in Equation \ref{externaltorque} and that on short timescales it changes only due to the varying $\Omega_{\rm c}(t)$, then
\beq
\frac{I_{\rm n}}{I_{\rm c}}\geq\frac{\dot{\Omega}^{\rm post}_{\rm c}-\dot{\Omega}_{\rm c}^{\rm pre}}{\dot{\Omega}_{\rm c}^{\rm pre}} \, ,
\eeq 
where $\dot{\Omega}_{\rm c}^{\rm pre}$ is the spin-down rate immediately before the glitch and $\dot{\Omega}^{\rm post}_{\rm c}$ is at time $t>t_{\rm g}$ such that $\Omega_{\rm c}^{\rm post}<\Omega_{\rm c}^{\rm pre}$. 
We evaluate this using the epochs of the TOAs surrounding each glitch as $t_{\rm pre}$ and $t_{\rm post}$. 
Although an earlier $t_{\rm post}$ would lead to tighter constraints, we prefer to avoid a less accurate extrapolation of the timing model closer to the (unknown) glitch epoch. 
The obtained constraint is \[\frac{I_{\rm n}}{I_{\rm c}}\geq1.4\times10^{-3}\;\text{.}\] 

While observed $\Delta\dot{\Omega}_{\rm c}/\dot{\Omega}_{\rm c}$ is usually $10^{-4}-10^{-3}$, immediately after the glitch the decoupled fraction of the superfluid can be much higher, as has been inferred for other pulsars \citep{lss00,dcl02}. Usually such large changes recover quickly, often in an exponential manner with various characteristic timescales, from few hours to several days.  
If present, such a strong relaxation would have been easily missed for PSR~J0537$-$6910: its very fast rotation compared to other pulsars shortens considerably the re-coupling timescales, while the monitoring cadence was not very high. 
Both glitch parameters $\Delta\Omega$ and $\Delta\dot{\Omega}$ should then be viewed as lower limits of their counterparts at the glitch epoch. 

\subsection{Comparison to other pulsars}

PSR~J0537$-$6910 is often compared to another very energetic pulsar located in the Large Magellanic Cloud, J0540$-$6919, and to the Crab pulsar (J0534$+$2200). These three neutron stars have the highest $|\dot{\nu}|$ and $\dot{E}$ amongst all rotation-powered pulsars, but their glitch activities are very different. Only two small glitches have been observed in PSR~J0540-6919 in $\sim16$ years \citep{fak15}, while the Crab had 24 glitches in 45 years, of small to intermediate sizes that follow a power-law distribution \citep{eas+14}. Due to this low activity, neither their glitch rate nor their average spin increase per time due to glitches follow the general positive correlation with $|\dot{\nu}|$ observed in glitching pulsars, while PSR~J0537$-$6910 just about does \citep{els+11,fes17}. The cumulative glitch increase of $\nu$ over time for the Crab is far from linear, unlike that of PSR~J0537$-$6910 (Figure \ref{fig:glitches}). These differences can be attributed to the age of J0537$-$6910 ($\gtrsim 4$ times older than the Crab according to the supernova remnant estimates and the -- less reliable -- characteristic $\tau_{\rm sd}$) and/or to a likely smaller internal temperature, if a connected glitch-driving region $I_{\rm G}$ able to sustain a large $\omega$ is not yet formed in the other two pulsars. Their glitches could then be due to local vortex avalanches that do not have a natural length- and time-scale \citep[see for example ][]{h16}. Another notable difference with J0537$-$6910 is that the Crab displays a low inter-glitch braking index ($n_{\rm ig}\simeq2.5$), which is very close to its long-term value of $2.3$ \citep{ljgesw15}.  

In fact, the glitch activity of J0537$-$6910 resembles much more that of the Vela pulsar and the Vela-like pulsars J1803-2137 and J1826-1334. Glitches in Vela have usually large $\Delta\nu$, close to normally distributed, and are rather regular (every $\sim3-3.5$ years). The cumulative glitch increase of $\nu$ over time is also linear, with an inferred $I_{\rm G}$ of about 1.62(3)\% \citep{hea+15}. The much smaller glitch rate of Vela ($\sim0.5$ events per year) compared to that of J0537$-$6910 can be explained by the difference in their spin-down rate: the increase in the lag $\omega_{\rm G}$ between Vela glitches is $\sim3$ times smaller than that for J0537$-$6910. We do not see a correlation between glitch size and the time to the next glitch in Vela. However, its glitch sample found in the literature is not complete - small glitches often stay unreported. Although these small events have little impact on the measurement of $I_{\rm G}$, they affect the waiting times distribution. Besides, the presence of strong exponentially relaxing components perplexes the measurements of $\Delta\nu$ in a consistent way for all Vela glitches. Similarly to PSR~J0537$-$6910, the average inter-glitch braking index for Vela is very high ($\sim40$) while the long-term one is rather low ($=1.7\pm0.2$); this is also the case for the other two Vela-like pulsars \citep{els17}. 


 \section{The long-term evolution of the spin-down rate}

Opposite to the inter-glitch behaviour, which is characterised by a large, positive $\ddot{\nu}$, the overall change in $\dot{\nu}$ is negative and corresponds to a decrease of $\sim 3.2\times10^{-13} \;\rm{Hz\,s^{-1}}$ over the $\sim13$ years of data (see Figure \ref{fig:F1}). 
This implies a negative long-term braking index and decreasing characteristic age, as already noted by \citet{mgm+04}. 
Phase-coherent timing and a direct fit for the braking index of a power-law spin-down model is not possible due to the frequent glitches. A simple quadratic fit, however,  of derived spin frequency data over the entire timespan leads to $\nu\ddot{\nu}/\dot{\nu}^2=-1.2(1)$. 
This method is inaccurate, mostly due to the presence of glitches, but demonstrates the clear detection of an enhancement of the spin-down rate in time.   
We use two ways to quantify this behaviour in terms of a long-term ``braking index'' $n_{\rm l}$.  
Firstly, we perform a linear fit to two different sets of $\dot{\nu}$ data. 
One set consists of the measurements $\dot{\nu}_0$ for the fit epoch of the inter-glitch timing solutions (Table \ref{tbl:f2s}), while for the second set we use $\dot{\nu}$ values extrapolated at an epoch immediately before the next glitch. 
The results of the fit are consistent with each other and give $\ddot{\nu}=-7.6(7)\times10^{-22}\;\rm{Hz\,s^{-2}}$. 
Secondly, we use the techniques described in \citet{els17} for measuring braking indices in pulsars with large and regular glitches. The method is based on constructing a $\dot{\nu}$-template which is used to calculate the relative $\dot{\nu}$ shifts of each inter-glitch interval. Together with the glitch epochs, these define a set of $(\dot{\nu},t)$  points that follow a straight line (linear correlation coefficient $0.976$) from which the long-term $\ddot{\nu}$ can be inferred. For PSR~J0537$-$6910, we use the 39 largest glitches and a 160-days long template constructed from their post-glitch curves. In line with our previous estimates, we obtain a long-term $\ddot{\nu}=-7.7(3)\times10^{-22}\;\rm{Hz\,s^{-2}}$ and a corresponding $n_{\rm l}=-1.22(4)$. 

Identifying the possible mechanism behind the unusual increase in spin-down rate depends on whether this change is continuous, or happens in discrete steps (for example in relation to the glitches), or a combination of both. Unfortunately we were not able to confidently discriminate between these options from the data at hand. 
We outline some possibilities below and give quantitative estimates under the assumption that - for the time interval of interest - the actual braking index $n$ is constant. To obtain numerical results we often assume an underlying power-law braking with $n=3$, however various mechanisms contribute to the evolution of $\dot{\nu}$, some of which might not even have a power-law form (one such example could be the internal torque due to the superfluid dynamics). Moreover, there is a wide range of reported long-term braking indices for other young pulsars, the majority of which is under 3 (see for example \citet{ljgesw15,els17} and references therein). Although some of the same physical processes might be responsible for all observed low ($<3$) braking indices, PSR~J0537$-$6910 is the only rotationally-powered pulsar with a long-term $\ddot{\nu}<0$ that is well-defined over so many years. 

A decreasing effective moment of inertia $I_{\rm c}$ could result in braking indices less than three. 
\citet{ha12} modelled a decreasing $I_{\rm c}$ as the result of an ongoing formation of new superfluid regions as they cool below the critical temperature for superfluidity. An increasing fraction of the interior then decouples, leaving a smaller moment of inertia to respond to the (nearly constant) spin-down torque (which we take here as the standard dipole braking, Eq. \ref{externaltorque}). If the two fluids are completely decoupled ($\dot{\Omega}_{\rm s}=0$) and $(\Omega_{\rm s}-\Omega_{\rm c})/\Omega_{\rm c}\ll1$ then 
\[
n_{\rm l} \simeq 3 -2 \frac{\dot{I}}{I}\frac{\Omega_{\rm c}}{\dot{\Omega}_{\rm c}} \; \Rightarrow \; \dot{I}\simeq2\times10^{41}\left(\frac{I}{10^{45}\rm{g\,cm^2}}\right) \; \rm{g\,cm^2\,yr^{-1}} \]
is required to explain the observed braking index of $-1.2$. One of the main challenges of this model is to accommodate the low braking index of the - relatively old and cool - Vela pulsar consistently with the constraints for the (de)coupled superfluid moment of inertia that come from glitch observations\footnote{$\dot{\Omega}_{\rm s}=0$ cannot hold in the bulk of the star if the post-glitch relaxation is of superfluid origin.}. 
A similar, also short-lived compared to $\tau_{\rm sd}$, effect of decreasing $I_{\rm c}$ will arise if the coupling in some, already superfluid, regions becomes weaker; this can be done also in discrete steps, for example if small crustquakes leave the nuclei lattice deformed in such a way that vortex pinning becomes stronger \citep{acc+96}.

Accumulated offsets $\Delta\dot{\nu}<0$, due to the frequent glitches which decouple the internal superfluid, could mimic a negative long-term $\ddot{\nu}$. 
After all, the very high $n_{\rm ig}$ up to the time of a following glitch means that the recoupling of the superfluid still carries on. 
This mechanism requires the progressive decoupling of a superfluid effective moment of inertia $I_{\rm s}$ additional to $I_{\rm n}$ that does not recouple on observable timescales. 
If the persisting offsets arose only from parts of the $I_{\rm n}$ component that did not have the time to recover till the next glitch, then $|\dot{\Omega}_{\rm c}|^{\rm min}$ would change but $|\dot{\Omega}_{\rm c}|^{\rm max}$ would remain limited at $|N_{\rm ext}|/I_{\rm c}$. Instead, as clearly seen in Figure \ref{fig:F1}, $|\dot{\Omega}_{\rm c}|^{\rm max}$ also follows an increasing trend; moreover, measurements of $\Delta\dot{\nu}/\dot{\nu}$ show no convincing evidence for a decreasing $I_{\rm n}/I_{\rm c}$ over time. In this scenario, the $I_{\rm s}$ that needs to have decoupled during the {\it RXTE} observations needs to be of the order $10^{-3}I$. 

It is worth noting that glitches in at least two pulsars left long-lasting decreases in their braking indices \citep{lnk+11,awe+15}, mainly due to a drop in $\ddot{\nu}$ after the initial post-glitch recovery stages were over. A period of higher glitch activity of the Crab pulsar is also associated with a decrease of its braking index, even after some persisting glitch offsets have been corrected for \citep{ljgesw15}, and 9 out of the 12 pulsars with a measured long-term $n<3$ have known glitches \citep[][and references therein]{els17}. It is perhaps then not surprising that PSR~J0537$-$6910, with its unprecedented high rate of large glitches, has the smallest braking index of all. 

Another possibility is an increasing torque as a result of magnetic field evolution. 
We consider a spin-down that is governed by \[ \dot{\Omega}_{\rm c}=-f(\star)B_\star^2\Omega_{\rm c}^n \] which makes the assumption that the superfluid is either completely decoupled ($\dot{\Omega}_{\rm s}=0$) or in equilibrium with the normal component ($\dot{\Omega}_{\rm s}=\dot{\Omega}_{\rm c}$). 
Here, $f(\star)$ is a function of stellar parameters that will be taken constant in time, $B_\star=g(\alpha) B_{\rm p} $ with $g(\alpha)$ a trigonometric function of $\alpha$, and $n$ is the real, underlying braking index. 
Allowing for a changing $B_\star$ means that the observed braking index will be 
\beq
\label{nlB}
n_{\rm l}=n+2\frac{\Omega_{\rm c}}{\dot{\Omega}_{\rm c}}\frac{\dot{B}_\star}{B_\star}
\eeq
which implies, for $n_{\rm l}=-1.2$, $n=3$ and constant $g(\alpha)$:  
\beq
 \dot B_{\rm p}\simeq6.8\left(\frac{|\dot{\nu}_{\rm c}|}{2\times10^{-10} \rm{Hz\,s^{-1}}}\right)\left(\frac{\nu}{62\rm{Hz}}\right)^{-1}\left(\frac{B_{\rm p}}{10^{12}\rm{G}}\right) \;\rm{G\,s^{-1}\,.}
 \eeq
It has been suggested that such a $\dot B_{\rm p}$ could be due to field reemergence from the crust following an initial mass accretion \citep[see for example][]{mp96,pvg12}, although it is unclear whether PSR~J0537$-$6910 could have accreted enough material for the required field burial because it has a large inferred kick velocity (\citealt{nr07}, but see also \citealt{ge13}). An amplification of the dipole component due to Hall drift at early timescales has been also proposed \citep{gc15}, but requires an exceptionally strong magnetic field in the crust.

Alternatively, $\dot{B}_\star$ can be due to an increasing misalignment of the magnetic and rotational axes, while the magnetic field strength remains almost constant ($\dot{B}_{\rm p}=0$). 
\citet{mmw+06} explored this idea in detail, taking $\alpha$ to be a linear function of time and the intrinsic braking index $n$ to be less than $3$, to track the birth values of $\alpha$ and $\nu$.  
Discrete shifts  $\Delta\alpha$ in the inclination angle have been invoked to explain persisting changes of the rotational parameters after glitches, seen in several pulsars \citep{leb92,le97,aga15}. In the case of the Crab pulsar, however, the sum of possible glitch-associated $\Delta\alpha$ over the total observing time does not suffice to explain the deviation of its braking index from 3 \citep{ljgesw15}, nor the recently reported evidence for increasing phase separation between its main pulse and interpulse \citep{lgwjsbk13} which suggests a non-zero inter-glitch $\dot{\alpha}$. 
From Equation \ref{nlB}, for $g(\alpha)=\sin\alpha(t)$, $B_{\rm p}=\rm{const}$ we get 
\(
\cot\alpha\,\dot{\alpha}=(n_{\rm l}-n)\dot{\nu}/(2\nu) 
\)
or, for $n=3$ and $n_{\rm l}=-1.2$:
\beq
\frac{\dot{\alpha}}{\tan\alpha}=6.8\times10^{-12}\left(\frac{|\dot{\nu}|}{2\times10^{-10}\rm{Hz\,s^{-1}}}\right)\left(\frac{\nu}{62\rm{Hz}}\right)^{-1} \;\rm{rad\,s^{-1}}\, .
\eeq
For a moderate $\alpha=30^\circ$ we get $\dot{\alpha}\simeq0.7$ degrees per century, remarkably close to the slow migration rate inferred for the Crab pulsar \citep{lgwjsbk13}.
  
\section{Summary and conclusions}

We analysed all {\it RXTE} observations of the pulsar J0537$-$6910 with the aim of reconstructing its rotational history and studying its glitching activity.
Twenty-one new glitches were identified, raising their total number for this pulsar at 45 in nearly 13 years. 
Most spin-ups are large and follow a broad Gaussian distribution with mean $\Delta\nu\sim16 \,\rm{\mu Hz}$. 
The first glitch in the data remains the largest and the only one for which some exponential decay of the spin-up is seen, on a characteristic timescale of $\sim20$ days. The rest were adequately described by a model with step-like changes in the spin $\nu$, spin-down $\dot{\nu}$ and sometimes $\ddot{\nu}$.  
The mean waiting time between glitches is $105.5$ days and is strongly correlated to the size $\Delta\nu$ of the preceding glitch. 
This fact can be used to predict the epoch of the subsequent spin-up. 

Almost all glitches display an abrupt spin-down rate change, $\Delta\dot{\nu}<0$, with mean $\Delta\dot{\nu}\sim-120\times10^{-12}\,\rm{Hz\,s^{-1}}$. 
A correlation between the waiting time and the size $|\Delta\dot{\nu}|$ of the following glitch, with a saturation at $|\Delta\dot{\nu}|\simeq150\times10^{-12}\,\rm{Hz\,s^{-1}}$ was noticed by \citet{mmw+06}. Such a finding would provide extra support for superfluid models of post-glitch relaxation but unfortunately its concrete confirmation was not possible using our results. 

In a simple 3-component model of a neutron star, the observed glitch phenomenology can be explained as follows: A part of the superfluid, bearing less than $0.8\%$ of the total moment of inertia, does not spin down at all in between glitches. Once its rotational velocity exceeds that of the crust by $\gtrsim0.03\,\rm{rad\,s^{-1}}$, it rapidly transfers some angular momentum to the non-superfluid (normal) component, causing a glitch. The rest of the superfluid follows the spin-up of the normal component on timescales that depend on coupling strength between the two; an effective $\gtrsim0.15\%$ of the total moment of inertia remains decoupled on long enough timescales so that the observed spin-down rate is higher post-glitch. The slow recoupling of this component is most likely responsible for the large inter-glitch braking indices ($\overline{n}_{\rm ig}\sim22$). 

Over the time span of the observations, the spin-down rate increased by $\sim0.15\%$, equivalent to a well-defined negative long-term $\ddot{\nu}=-7.5(3)\times10^{-22}\;\rm{Hz\,s^{-2}}$, unlike what is seen in any other young pulsars. The inferred braking index for this behaviour is $n_{\rm l}=-1.22(4)$ and could be attributed to the progressive decoupling of a superfluid component (of $\sim10^{-3}$ fractional moment of inertia), or an increasing external spin-down torque, e.g. as the pulsar becomes more of an orthogonal rotator. 
The pulsar's evolutionary path on the $P-\dot{P}$ diagram points in the direction of the Crab pulsar \citep{els17}. Clearly though, the duration of the phase with this low braking index cannot be very long compared to $\tau_{sd}$. Although we were unable to resolve a significant time variation of $n_{\rm l}$, it is not impossible --albeit optimistic-- to imagine that a measurable change could be revealed in the coming years, given a good observational coverage. 

PSR~J0537$-$6910 has the highest spin frequency and spin-down luminosity of all known rotational-powered pulsars and a remarkable rotational evolution, vastly dominated by large glitches. 
The high glitch rate of $\sim3.5$ per year and the predictability of the time for a next glitch, given a firm measurement of the preceding one, makes this pulsar an ideal target for close observations of a glitch's rise, early recovery and possible emission changes with missions like the {\it Neutron star Interior Composition Explorer (NICER)}, which are crucial for the advances of our theoretical understanding of the phenomenon. Furthermore, long-term monitoring might shed light to the mechanism behind the exceptional negative long-term braking index observed for this pulsar.

\section*{Acknowledgments}
D.A. acknowledges support from the Polish National Science Centre (SONATA BIS 2015/18/E/ST9/00577, P.I.: B.~Haskell) and the research networking programme NewCompStar (COST Action MP1304). 
C.M.E. acknowledges support from FONDECYT Regular Project 1171421.

\bibliographystyle{mnras}
\bibliography{journals,0537Bib}

\label{lastpage}
\end{document}